\documentclass[12pt]{article}

\usepackage{graphicx}
\begin{document}

\begin{center}
{\bf Magnetic black hole thermodynamics in an extended phase space with nonlinear electrodynamics} \\
\vspace{5mm} S. I. Kruglov
\footnote{E-mail: serguei.krouglov@utoronto.ca}
\underline{}
\vspace{3mm}

\textit{Department of Physics, University of Toronto, \\60 St. Georges St.,
Toronto, ON M5S 1A7, Canada\\
Canadian Quantum Research Center, \\
204-3002 32 Ave., Vernon, BC V1T 2L7, Canada} \\
\vspace{5mm}
\end{center}
\begin{abstract}
We study Einstein's gravity coupled to nonlinear electrodynamics with two parameters in Anti-de Sitter spacetime. Magnetically charged black holes in an extended phase space is investigated. We obtain the mass and metric functions, their asymptotic and corrections to the Reissner--Nordstr\"{o}m metric function when the cosmological constant vanishes. The first law of black hole thermodynamics in extended phase space is formulated and the magnetic potential and the thermodynamic conjugate to the coupling are obtained. We proved the generalized Smarr relation. The heat capacity and the Gibbs free energy are computed and phase transitions are studied. It was shown that the electric field of charged objects at the origin and electrostatic self-energy are finite within the nonlinear electrodynamics proposed.
\end{abstract}


\section{Introduction}

It was understood that black hole area has the role of the entropy and the temperature is connected with surface gravity \cite{Bekenstein,Hawking,Bardeen,Jacobson,Padmanabhan}. The importance of gravity in AdS space-time is due to the holographic
principle (a gauge duality description) \cite{Maldacena} which has applications in condensed matter physics. Firstly black hole phase transitions in Schwarzschild-AdS spacetime were studied in Ref. \cite{Page}. The negative cosmological constant, in an extended phase space black hole thermodynamics, is linked with a thermodynamic pressure conjugated to the volume \cite{Dolan,Kubiznak,Teo,Mann1}. The cosmological constant variation was included in the first law of black hole thermodynamics in Refs. \cite{Caldarelli,Kastor,Dolan1,Dolan2,Dolan3,Cvetic2}. But within general relativity the cosmological constant $\Lambda$ is a fixed external parameter. Also, such variation of $\Lambda$ in the first law of black hole thermodynamics means consideration of black hole ensembles possessing different asymptotics. This point of view is different from standard black hole thermodynamics, where parameters are varied but AdS background is fixed. There are some reasons to consider the variation of $\Lambda$ in black hole thermodynamics. First of all physical constants can arise as vacuum expectation values and not fixed a priori, and therefore, may
vary. As a result, these ‘constants’ are not real constants and may be included in the first law of black hole thermodynamics \cite{Gibbons,Creighton}. The second reason is that without varying the cosmological constant the Smarr relation is inconsistent with the first law of black hole thermodynamics \cite{Kastor}. When $\Lambda$ is inserted in the first law of black hole thermodynamics, the mass M of black holes should be treated as enthalpy but not than internal energy \cite{Kastor}. The first law of black hole thermodynamics can be formulated within Einstein's gravity if one includes the $V dP$ term. This requires to introduce a negative cosmological constant $\Lambda$ as a
positive pressure $P=-\Lambda/(8\pi)$. As a result, we come to AdS space-time. It should be noted that thermodynamic pressure $P$ is different from the local pressure which is present in the energy-momentum tensor. The conjugate variable to $P$ is the thermodynamic volume $V = 4\pi r_+^3/3$, where $r_+$ is the event horizon radius of a black hole.

In this paper, we study Einstein-AdS gravity coupled to nonlinear electrodynamics (NED) with two parameters, proposed here, that allows us to smooth out singularities. First NED was Born--Infeld electrodynamics \cite{Born} without singularities of point-like particles possessing the electric self-energy finite and at weak-field limit it is converted into Maxwell’s theory. Our NED model has similar behaviour. We will study magnetic black holes and their thermodynamics in the Einstein-AdS gravity in the extended phase space. The NED model, with coupling $\beta$ and dimensionless parameter $\sigma$ proposed here, is  of interest because it includes model \cite{Kr} for $\sigma=1$. This united approach allows us to find similarities and differences for different models.

The structure of the paper is as follows. In section 2 we find the mass and metric functions and their asymptotic. Corrections to the Reissner--Nordstr\"{o}m metric function are obtained when the cosmological constant is zero. We proof the first law of black hole thermodynamics in extended phase space and obtain the magnetic potential and the thermodynamic conjugate to the coupling. The generalized Smarr formula is proved. The Gibbs free energy is calculated and depicted for some parameters and phase transitions are studied in section 3. In Appendix A we obtain the electric field of charged objects and corrections to Coulomb's law. We show that the electrostatic self-energy of charged particles is finite in Appendix B. In Appendix C we obtain the metric which is a solution of the Einstein--Maxwell system.
Section 4 is a discussion of results obtained.

We use units with $c=\hbar=k_B=1$.

\section{Einstein-AdS black hole solution}

The action of Einstein's gravity in AdS spacetime is given by
\begin{equation}
I=\int d^{4}x\sqrt{-g}\left(\frac{R-2\Lambda}{16\pi G}+\mathcal{L}(\mathcal{F}) \right),
\label{1}
\end{equation}
where $G$ is the gravitation constant, $\Lambda=-3/l^2$ is the negative cosmological constant and $l$ is the AdS radius. We propose the NED Lagrangian as follows:
\begin{equation}
{\cal L}(\mathcal{F})=-\frac{{\cal F}}{4\pi\left(1+2\beta {\cal F}\right)^\sigma},
\label{2}
\end{equation}
where ${\cal F}=F^{\mu\nu}F_{\mu\nu}/4=(B^2-E^2)/2$ is the Lorenz invariant and $E$ and $B$ are the electric and magnetic fields, correspondingly. The coupling $\beta>0$ has the dimension $L^4$, and the dimensionless parameter $\sigma>0$.  The weak-field limit of Lagrangian (2) is the Maxwell's Lagrangian. The Lagrangian (2) at $\sigma=1$ becomes the rational NED Lagrangian \cite{Kruglov}.
The NED Lagrangian (2) for some values of $\sigma$ was used in the inflation scenario \cite{Kruglov1,Kruglov3,Kruglov4}.
From action (1) one finds the Einstein and field equations
\begin{equation}
R_{\mu\nu}-\frac{1}{2}g_{\mu \nu}R+\Lambda g_{\mu \nu} =8\pi G T_{\mu \nu},
\label{3}
 \end{equation}
\begin{equation}
\partial _{\mu }\left( \sqrt{-g}\mathcal{L}_{\mathcal{F}}F^{\mu \nu}\right)=0,
\label{4}
\end{equation}
where $\mathcal{L}_{\mathcal{F}}=\partial \mathcal{L}( \mathcal{F})/\partial \mathcal{F}$. The energy-stress tensor reads
\begin{equation}
 T_{\mu\nu }=F_{\mu\rho }F_{\nu }^{~\rho }\mathcal{L}_{\mathcal{F}}+g_{\mu \nu }\mathcal{L}\left( \mathcal{F}\right).
\label{5}
\end{equation}
We consider here spherical symmetry with the line element
\begin{equation}
ds^{2}=-f(r)dt^{2}+\frac{1}{f(r)}dr^{2}+r^{2}\left( d\theta
^{2}+\sin ^{2}(\theta) d\phi ^{2}\right).
\label{6}
\end{equation}
The magnetic black holes possess the magnetic field $B=q/r^2$ where $q$ is the magnetic charge.
The metric function is given by (see Appendix C and \cite{Bronnikov})
\begin{equation}
f(r)=1-\frac{2m(r)G}{r},
\label{7}
\end{equation}
with the mass function
\begin{equation}
m(r)=m_0+4\pi\int \rho (r)r^{2}dr,
\label{8}
\end{equation}
where $m_0$ is an integration constant (the Schwarzschild mass) and $\rho$ being the energy density.
We obtain the energy density
\begin{equation}
\rho=\rho_M-\frac{3}{8\pi Gl^2},
\label{9}
\end{equation}
where the magnetic energy density found from Eqs. (2) and (5) is
\[
\rho_M=\frac{q^2r^{4(\sigma-1)}}{8\pi \left(r^4+q^2\beta\right)^{\sigma}}.
\]
Making use of Eqs. (8) and (9) we obtain the mass function
\begin{equation}
m(r)=m_0+\frac{q^2r^{4\sigma-1}}{2(4\sigma-1)(q^2\beta)^\sigma} F\left(\sigma-\frac{1}{4},\sigma;\sigma+\frac{3}{4};-\frac{r^4}{q^2\beta}\right)-\frac{r^3}{2Gl^2},
\label{10}
\end{equation}
where $F(a,b;c;z)$ is the hypergeometric function. The magnetic energy is given by
\begin{equation}
m_M=4\pi\int_0^\infty\rho_M (r)r^{2}dr=\frac{q^{3/2}\Gamma(\sigma-1/4)\Gamma(5/4)}{2\beta^{1/4}\Gamma(\sigma)},
\label{11}
\end{equation}
where $\Gamma(x)$ is Gamma-function. Equation (11) shows that at the Maxwell's limit $\beta\rightarrow 0$ the black hole magnetic mass diverges.
Therefore, a smooth limit to Maxwell's theory is questionable. From Eqs. (7) and (10) one finds the metric function
\begin{equation}
f(r)=1-\frac{2m_0 G_N}{r}-\frac{q^2Gr^{4\sigma-2}}{(4\sigma-1)(q^2\beta)^\sigma} F\left(\sigma-\frac{1}{4},\sigma;\sigma+\frac{3}{4};-\frac{r^4}{q^2\beta}\right)+\frac{r^2}{l^2}.
\label{12}
\end{equation}
We use the relation \cite{Abramowitz}
\begin{equation}
F(a,b;c;z)=1+\frac{ab}{c}z+\frac{a(a+1)b(b+1)}{c(c+1)}z^2+....,
\label{13}
\end{equation}
for $|z|<1$, which will be used to obtain the asymptotic of the metric function as $r\rightarrow 0$.
When the Schwarzschild mass is zero ($m_0=0$) and as $r\rightarrow 0$, the asymptotic  is
\begin{equation}
f(r)=1+\frac{r^2}{l^2}-\frac{Gq^2r^{4\sigma-2}}{(q^2\beta)^\sigma(4\sigma-1)}+\frac{G\sigma r^{4\sigma+2}}{\beta^{\sigma+1}q^{2\sigma}(4\sigma+3)}+{\cal O}(r^{4\sigma+6}).
\label{14}
\end{equation}
Equation (14) shows that at $\sigma\geq 1/2$ a singularity of the metric function $f(r)$ is absent. In addition, to avoid  a conical singularity at $r=0$  we also should set $4\sigma-2> 1$ ($\sigma>3/4$). It worth noting that the magnetic
energy density $\rho_M$ is finite at $r=0$ only if $\sigma\geq 1$. Therefore, to have regular black holes one has to assume that $\sigma\geq 1$. Then from Eq. (14) we find $f(0)=1$ that is a necessary condition to have the spacetime regular.
We explore the transformation \cite{Abramowitz}
\[
F(a,b;c;z)=\frac{\Gamma(c)\Gamma(b-a)}{\Gamma(b)\Gamma(c-a)}(-z)^{-a}F\left(a,1-c+a;1-b+a;\frac{1}{z}\right)
\]
\begin{equation}
+\frac{\Gamma(c)\Gamma(a-b)}{\Gamma(a)\Gamma(c-b)}(-z)^{-b}F\left(b,1-c+b;1-a+b;\frac{1}{z}\right),
\label{15}
\end{equation}
to obtain the asymptotic of the metric function as $r\rightarrow \infty$. By virtue of Eqs. (13) and (15) we find
\begin{equation}
f(r)=1-\frac{2(m_0+m_M) G}{r}+\frac{q^2G}{r^2}F\left(\sigma,\frac{1}{4};\frac{5}{4};-\frac{q^2\beta}{r^4}\right) +\frac{r^2}{l^2},
\label{16}
\end{equation}
where the relation $\Gamma(1+z)=z\Gamma(z)$ and $F\left(a,0;c;z\right)=1$ were used. Making use of Eqs. (13) and (16) as $r\rightarrow\infty$ when the cosmological constant vanishes ($l\rightarrow\infty$) we find
\begin{equation}
f(r)=1-\frac{2M G}{r}+\frac{q^2G}{r^2}-\frac{q^4\beta\sigma G}{5r^6}+{\cal O}(r^{-10}),
\label{17}
\end{equation}
where $M=m_0+m_M$ is the ADM mass (the total black hole mass as $r\rightarrow\infty$). It follows from Eq. (17) that corrections to the Reissner--Nordstr\"{o}m solution are in the order as $\mathcal{O}(r^{-6})$.
When $\beta\rightarrow 0$ the metric function (17) is converted into the Reissner--Nordstr\"{o}m metric function. The plot of metric function (12) is given in Fig. 1 with $m_0=0$, $G=q=1$, $\beta=0.1$, $l=5$.
\begin{figure}[h]
\includegraphics [height=3.0in,width=4.0in] {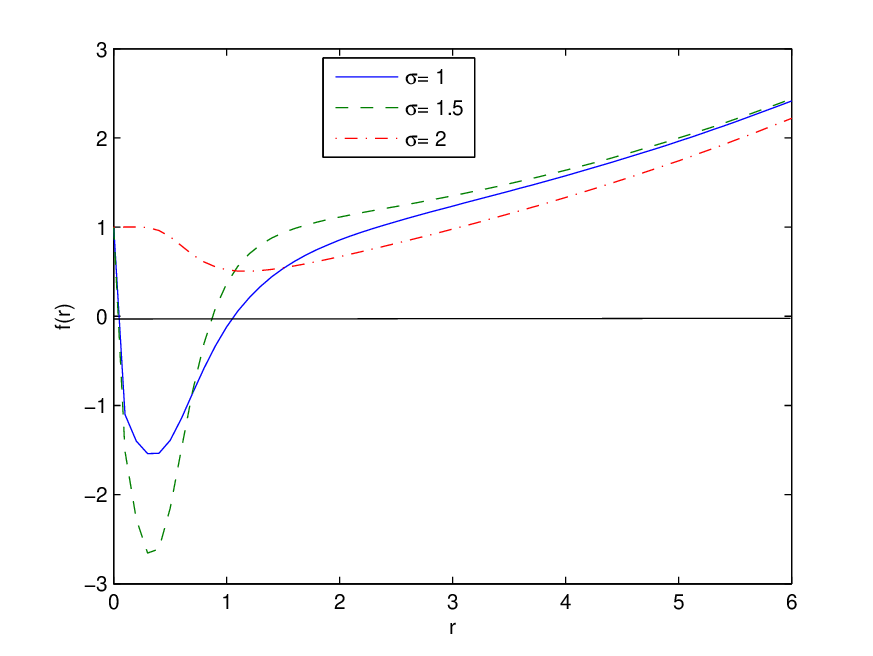}
\caption{\label{fig.1} The function $f(r)$ at $m_0=0$, $G=1$, $q=1$, $\beta=0.1$, $l=5$. Figure 1 shows that black holes may have one or two horizons. When $\sigma$ increases the event horizon radius $r_+$ decreases.}
\end{figure}
In accordance with Fig. (1), if parameter $\sigma$ increases the event horizon radius $r_+$ decreases. Figure 1 shows that black holes can have one or two horizons. It should be noted that when we set $G=c=\hbar=1$ as in Fig. 1, we come to Planckian units \cite{Mukhanov}. Then in this case if one has, for example, dimensionless event horizon radius $r_+=1$ (as in Fig. 1), in usual units $r_+=l_{Pl}=(G\hbar/c^3)^{1/2}=1.616\times 10^{-33}$ cm, where $l_{Pl}$ is Planck's length. When the dimensionless mass is $m=2$, for example, in
usual units $m=2\times m_{Pl}=2\times(\hbar c/G)^{1/2}= 2\times 2.177\times 10^{-5}$ g, where $m_{Pl}$ is Planck's  mass. Because in Fig. 1 the event horizon radius is small we have here the example of  tiny black holes (primordial black holes). Such  black holes could be just created after Big Bang. It is worth noting that such an example of quantum-sized black holes are described here by semiclassical gravity.

\section{First law of black hole thermodynamics}

The pressure, in extended phase space thermodynamics, is defined as $P=-\Lambda/(8\pi)$  \cite{Kastor,Dolan1,Cvetic2,Kubiznak2}. The coupling $\beta$ is treated as the thermodynamic value and the mass $M$ is a chemical enthalpy so that $M=U+PV$ and $U$ being the internal energy.
In the following we will use Planckian units with $G=c=\hbar=1$.
By using the Euler's dimensional analysis \cite{Smarr}, \cite{Kastor},
we have dimensions $[M]=L$, $[S]=L^2$, $[P]=L^{-2}$, $[J]=L^2$, $[q]=L$, $[\beta]=L^2$ and
\begin{equation}
M=2S\frac{\partial M}{\partial S}-2P\frac{\partial M}{\partial P}+2J\frac{\partial M}{\partial J}+q\frac{\partial M}{\partial q}+2\beta\frac{\partial M}{\partial \beta},
\label{18}
\end{equation}
where $J$ is the black hole angular momentum. In the following we consider non-rotating black holes and, therefore, $J=0$.
The thermodynamic conjugate to coupling $\beta$ is ${\cal B}=\partial M/\partial \beta $ (so-called vacuum polarization) \cite{Teo}. The black hole volume $V$, entropy $S$ are defined as
\begin{equation}
V=\frac{4}{3}\pi r_+^3,~~~~~~S=\pi r_+^2.
\label{19}
\end{equation}
From Eq. (16), and equation $f(r_+)=0$, where $r_+$ is the event horizon radius, one finds
\begin{equation}
M(r_+)=\frac{r_+}{2}+\frac{q^2}{2r_+}F\left(\sigma,\frac{1}{4};\frac{5}{4};-\frac{q^2\beta}{r_+^4}\right) +\frac{r_+^3}{2l^2}.
\label{20}
\end{equation}
Making use of Eq. (20), we obtain
\[
dM(r_+)=\biggl[\frac{1}{2}+\frac{3r_+^2}{2l^2}- \frac{q^2}{2r_+^2} F\left(\sigma,\frac{1}{4};\frac{5}{4};-\frac{q^2\beta}{r_+^4}\right)
\]
\[
+\frac{2\sigma q^4\beta}{5r_+^6} F\left(\sigma+1,\frac{5}{4};\frac{9}{4};-\frac{q^2\beta}{r_+^4}\right)\biggr]dr_+
-\frac{r_+^3}{l^3}dl
\]
\[
+\biggl[\frac{q}{r_+}
F\left(\sigma,\frac{1}{4};\frac{5}{4};-\frac{q^2\beta}{r_+^4}\right)
-\frac{q^3\beta\sigma}{5r_+^5}F\left(\sigma+1,\frac{5}{4};\frac{9}{4};-\frac{q^2\beta}{r_+^4}\right)\biggr]dq
\]
\begin{equation}
-\frac{q^4\sigma}{10r_+^5} F\left(\sigma+1,\frac{5}{4};\frac{9}{4};-\frac{q^2\beta}{r_+^4}\right)d\beta.
\label{21}
\end{equation}
Here, we have used the relation \cite{Abramowitz}
\begin{equation}
\frac{dF(a,b;c;z)}{dz}=\frac{ab}{c}F(a+1,b+1;c+1;z).
\label{22}
\end{equation}
Defining the Hawking temperature
\begin{equation}
T=\frac{f'(r)|_{r=r_+}}{4\pi},
\label{23}
\end{equation}
where $f'(r)=\partial f(r)/\partial r$, and by virtue of Eqs. (16) and (23), we obtain
\begin{equation}
T=\frac{1}{4\pi}\biggl[\frac{1}{r_+}+\frac{3r_+}{l^2}- \frac{q^2}{r_+^3} F\left(\sigma,\frac{1}{4};\frac{5}{4};-\frac{q^2\beta}{r_+^4}\right)
+\frac{4\sigma q^4\beta}{5r_+^7} F\left(\sigma+1,\frac{5}{4};\frac{9}{4};-\frac{q^2\beta}{r_+^4}\right)\biggr].
\label{24}
\end{equation}
At $\beta=0$ in Eq. (24), one finds the  Maxwell-AdS black hole Hawking temperature.
The first law of black hole thermodynamics follows from Eqs. (19), (20) and (24),
\begin{equation}
dM = TdS + VdP + \Phi dq + {\cal B}d\beta.
\label{25}
\end{equation}
From Eqs. (21) and (25) we obtain the magnetic potential $\Phi$ and the thermodynamic conjugate to coupling $\beta$ (vacuum polarization) ${\cal B}$
\[
\Phi =\frac{q}{r_+}
F\left(\sigma,\frac{1}{4};\frac{5}{4};-\frac{q^2\beta}{r_+^4}\right)
-\frac{q^3\beta\sigma}{5r_+^5}F\left(\sigma+1,\frac{5}{4};\frac{9}{4};-\frac{q^2\beta}{r_+^4}\right),
\]
\begin{equation}
{\cal B}=-\frac{q^4\sigma}{10r_+^5} F\left(\sigma+1,\frac{5}{4};\frac{9}{4};-\frac{q^2\beta}{r_+^4}\right).
\label{26}
\end{equation}
The plots of $\Phi$ and ${\cal B}$ versus $r_+$ are depicted in Fig. 2.
\begin{figure}[h]
\includegraphics [height=3.0in,width=5.0in] {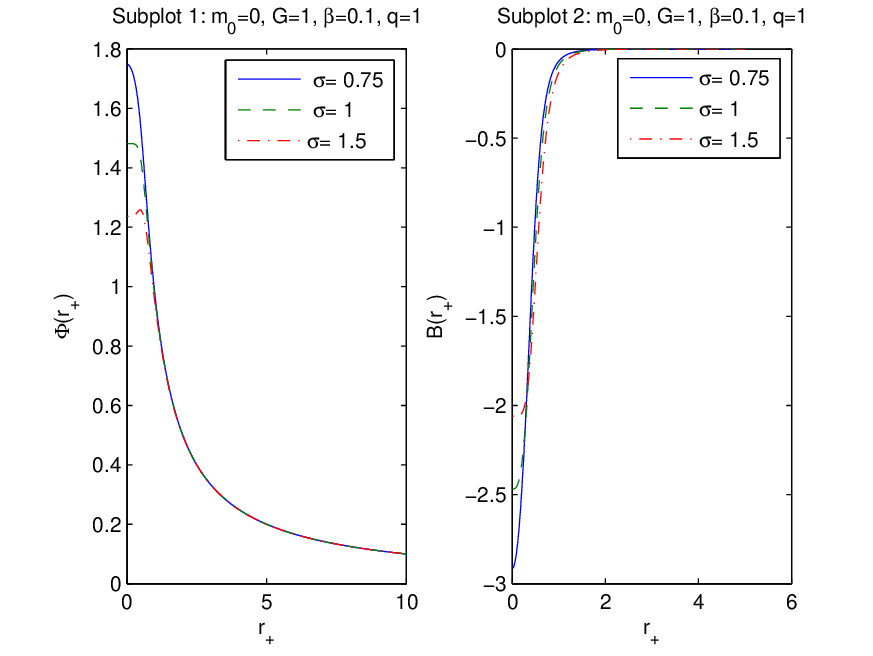}
\caption{\label{fig.2} The functions $\Phi$ and ${\cal B}$ vs. $r_+$ at $q=1$, $\beta=0.1$. The solid curve in subplot 1 is for $\sigma=0.75$, the dashed curve is for $\sigma=1$, and the dashed-doted curve is for $\sigma=1.5$. It follows that the magnetic potential $\Phi$ is finite at $r_+=0$ and becomes zero as $r_+\rightarrow \infty$. The function ${\cal B}$, in subplot 2, vanishes as $r_+\rightarrow \infty$ and is finite at $r_+=0$. }
\end{figure}
Figure 2, in the left panel, shows that as $r_+\rightarrow \infty$ the magnetic potential vanishes ($\Phi(\infty)=0$), and at $r_+ = 0$ $\Phi$ is finite. If the parameter $\sigma$ increases, $\Phi(0)$ decreases.
According to the right panel of Fig. 2 at $r_+ = 0$ the vacuum polarization is finite and as $r_+\rightarrow \infty$, ${\cal B}$ vanishes (${\cal B}(\infty)=0$). When the parameter $\sigma$ increases, ${\cal B}(0)$ also increases.
With the aid of Eqs. (19), (24) and (26) we find the generalized Smarr relation
\begin{equation}
M=2ST-2PV+q\Phi+2\beta{\cal B}.
\label{27}
\end{equation}

\section{Thermodynamics of black holes }

To study the local stability of black holes one can analyze the heat capacity
\begin{equation}
C_q=T\left(\frac{\partial S}{\partial T}\right)_q=\frac{T\partial S/\partial r_+}{\partial T/\partial r_+}=\frac{2\pi r_+ T}{\partial T/\partial r_+}.
\label{28}
\end{equation}
Equation (28) shows that when the Hawking temperature has an extremum the heat capacity possesses the singularity and the black hole phase transition occurs. With the help of Eq. (24) we depicted in Fig. (3) the Hawking temperature as a function of the event horizon radius.
\begin{figure}[h]
\includegraphics [height=3.0in,width=5.0in] {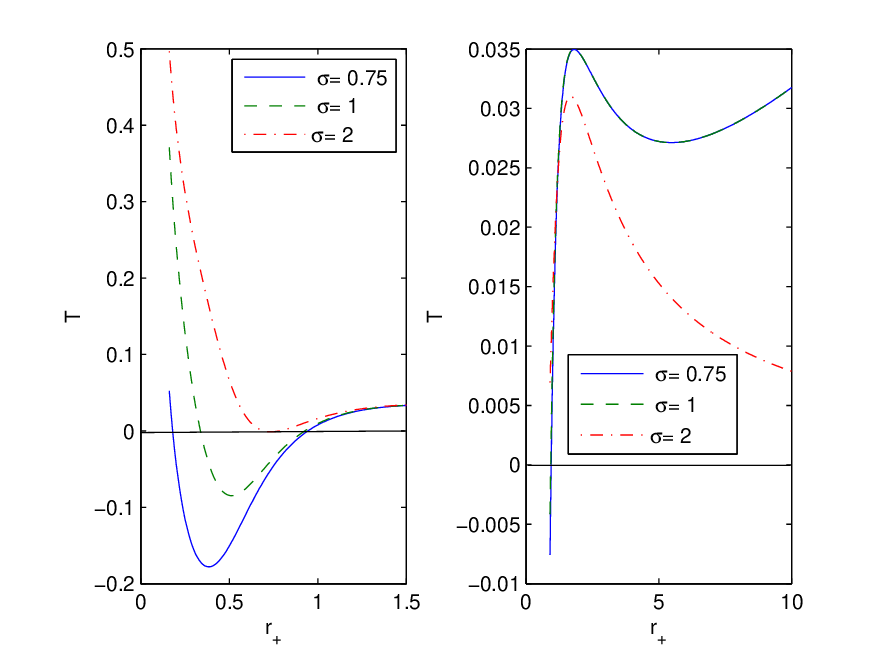}
\caption{\label{fig.3} The functions $T$ vs. $r_+$ at $q=1$, $\beta=0.1$, $l=10$. The solid curve in left panel is for $\sigma=0.75$, the dashed curve is for $\sigma=1$, and the dashed-doted curve is for $\sigma=2$. In some interval of $r_+$ the Hawking temperature is negative and, therefore, black holes do not exist at these parameters. There are extrema of the Hawking temperature $T$ where the black hole phase transitions occur.}
\end{figure}
For the case $\sigma=1$ the analysis of black holes local stability was performed in \cite{Kr}. The behavior of $T$ and $C_q$ depends on many parameters. By virtue of Eq. (24) we obtain
\[
\frac{\partial T}{\partial r_+}=\frac{1}{4\pi}\biggl[-\frac{1}{r_+^2}+\frac{3}{l^2}+ \frac{3q^2}{r_+^4} F\left(\sigma,\frac{1}{4};\frac{5}{4};-\frac{q^2\beta}{r_+^4}\right)
-\frac{32\sigma q^4\beta}{5r_+^8} F\left(\sigma+1,\frac{5}{4};\frac{9}{4};-\frac{q^2\beta}{r_+^4}\right)
\]
\begin{equation}
+\frac{16q^6\beta^2\sigma(4\sigma+1)}{9r_+^{12}} F\left(\sigma+2,\frac{9}{4};\frac{13}{4};-\frac{q^2\beta}{r_+^4}\right)\biggr].
\label{29}
\end{equation}
Equations (24) and (29) define the heat capacity (28). Making use of Eqs. (24), (28) and (29), one can study the heat capacity and the black hole phase transition for different parameters $\beta$, $\sigma$, $q$ and $l$.

With the help of Eq. (24) we obtain the black hole equation of state (EoS)
\begin{equation}
P=\frac{T}{2r_+}-\frac{1}{8\pi r_+^2}+\frac{q^2}{8\pi r_+^4}\biggl[F\left(\sigma,\frac{1}{4};\frac{5}{4};-\frac{q^2\beta}{r_+^4}\right)
- \frac{4q^2\beta\sigma}{5r_+^4}F\left(\sigma+1,\frac{5}{4};\frac{9}{4};-\frac{q^2\beta}{r_+^4}\right)\biggr].
\label{30}
\end{equation}
The specific volume is given by $v=2l_Pr_+$ ($l_P=\sqrt{G}=1$) \cite{Mann1}. Equation (30) is similar to EoS of the Van der Waals liquid. Putting $v=2r_+$ into expression (30) we obtain
\[
P=\frac{T}{v}-\frac{1}{2\pi v^2}+\frac{2q^2}{\pi v^4}\biggl[F\left(\sigma,\frac{1}{4};\frac{5}{4};-\frac{16q^2\beta}{v^4}\right)
\]
\begin{equation}
-\frac{64q^2\beta\sigma}{5v^4}F\left(\sigma+1,\frac{5}{4};\frac{9}{4};-\frac{16q^2\beta}{v^4}\right)\biggr].
\label{31}
\end{equation}
The plot of $P$ vs. $v$ is given in Fig. 4.
\begin{figure}[h]
\includegraphics [height=3.0in,width=4.0in] {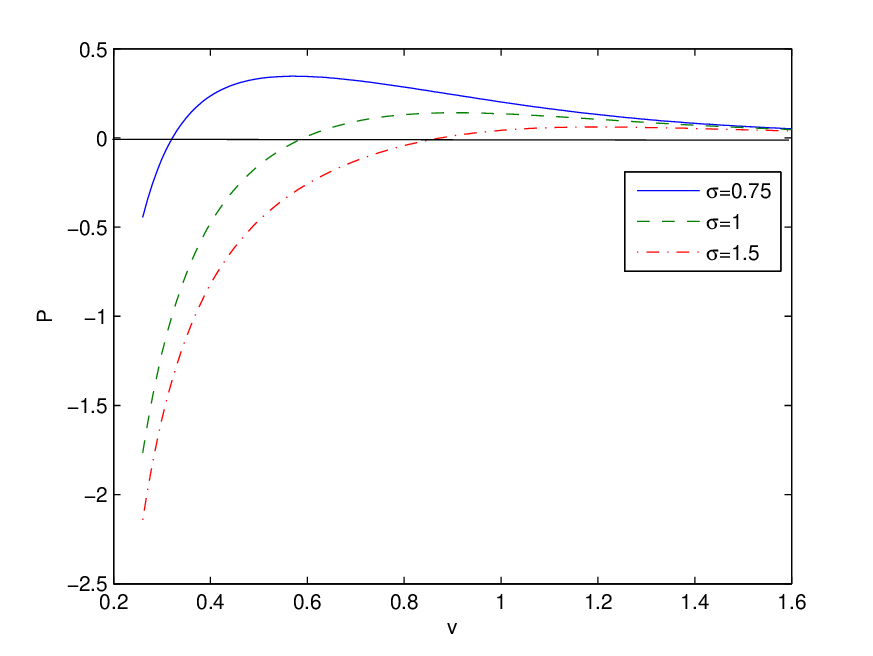}
\caption{\label{fig.4} The functions $P$ vs. $v$ at $q=1$, $\beta=0.1$, $T=0.05$. The solid line is for $\sigma=0.75$, the dashed curve is for $\sigma=1$, and the dashed-doted curve is for $\sigma=1.5$.}
\end{figure}
The critical points (inflection points) are defined by equations $\partial P/\partial v=0$, $\partial^2 P/\partial v^2=0$ which look cumbersome, so that we will not present them here. The analytical solutions for critical points do not exist. The $P-v$ diagrams at the critical values are similar to Van der Waals liquid diagrams having inflection points.

Because $M$ is treated as a chemical enthalpy the Gibbs free energy reads
\begin{equation}
G=M-TS.
\label{32}
\end{equation}
Making use of Eqs. (19), (20),(24) and (32) we obtain
\[
G=\frac{r_+}{4}-\frac{2\pi r_+^3P}{3}+\frac{3q^2}{4r_+}F\left(\sigma,\frac{1}{4};\frac{5}{4};-\frac{q^2\beta}{r_+^4}\right)
\]
\begin{equation}
-\frac{q^4\beta\sigma}{5r_+^5}F\left(\sigma+1,\frac{5}{4};\frac{9}{4};-\frac{q^2\beta}{r_+^4}\right).
\label{33}
\end{equation}
The plot of $G$ versus $T$ is given in Fig. 5 for $\beta=0.1$, $q=1$, $\sigma=0.75$.
\begin{figure}[h]
\includegraphics [height=3.0in,width=5.0in] {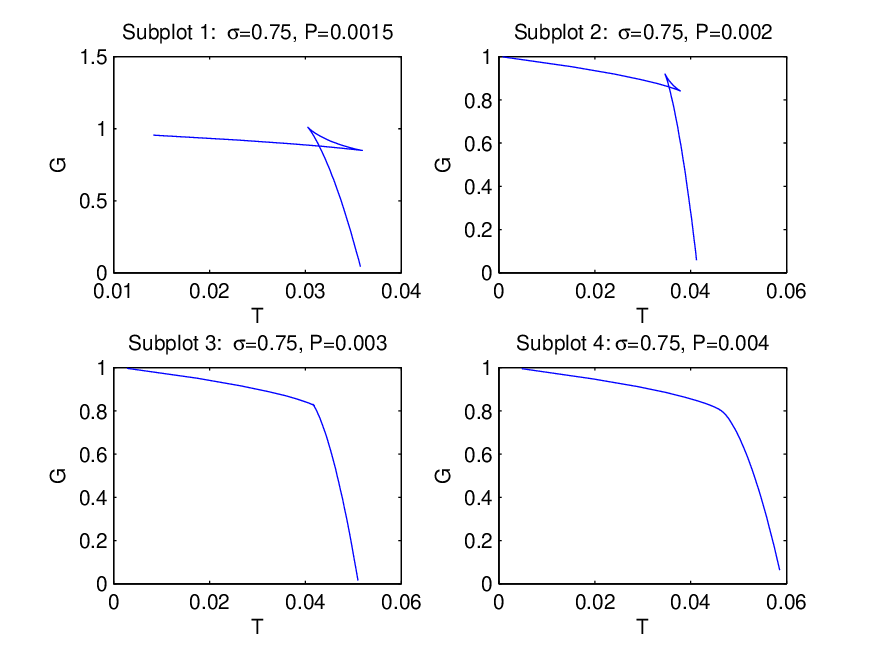}
\caption{\label{fig.5} The functions $G$ vs. $T$ at $q=1$, $\beta=0.1$, $\sigma=0.75$ for $P=0.0015$, $P=0.002$, $P=0.003$ and $P=0.004$.
Subplots 1 and 2 show the critical 'swallowtail' behavior with first-order phase transitions between small and large black holes. Subplots 3  corresponds to the case of critical point where second-order phase transition occurs ($P_c\approx 0.003$). Subplots 4 shows  non-critical behavior of the Gibbs free energy.}
\end{figure}
The critical points and phase transitions of black holes for $\sigma=1$ were studied in \cite{Kr}. One can investigate black holes phase transitions in our model for arbitrary $\sigma$ with the help of Gibbs's free energy (33).\\
It should be noted that analytical expressions obtained can be applied for black holes of any sizes. In Figures 1-5 we have
considered examples only for tiny black holes (quantum black holes).

\section{Summary}

We have obtained magnetic black hole solutions in Einstein-AdS gravity coupled to NED with two parameters which we propose here. The metric and mass functions and their asymptotic with corrections to the Reissner--Nordstr\"{o}m solution, when cosmological constant is zero, have been found. The total black hole mass includes the Schwarzschild mass and the magnetic mass that is finite. We have plotted the metric function showing that black holes may have one or two horizons. When parameter $\sigma$ increases the event horizon radius $r_+$ decreases. Figures 2-4 show how other physical variables depend on $\sigma$. The black holes thermodynamics in an extended phase space was studied. We formulated the first law of black hole thermodynamics where the pressure is connected with the negative cosmological constant (AdS spacetime) conjugated to the Newtonian geometric volume of the black hole. The thermodynamic potential conjugated to magnetic charge and the thermodynamic quantity conjugated to coupling $\beta$ (so called vacuum polarization), were computed and plotted. It was proofed that the generalized Smarr relation holds for any parameter $\sigma$. We calculated the Hawking temperature, the heat capacity and the Gibbs free energy. The analyses of first-order and second-order phase transitions were performed for some parameters. The Gibbs free energy has shown the critical 'swallowtail' behavior that is similar to the Van der Waals liquid–gas behavior. Figure 5 shown a first-order phase transition with Gibbs's free energy
which is continuous but not differentiable but for a second order transition the Gibbs free energy and its first derivatives are continuous. The same feature was firstly discovered for another model in \cite{Mann1}. It was shown within the NED proposed that the electric field of charged objects at the origin and electrostatic self-energy are finite. It should be noted that the first law of an electric black hole thermodynamics in the Einstein--Born--Infeld theory  and other problems were originally studied in Ref. \cite{Mann1}.\\

\textbf{Appendix A}\\

Making use of Eq. (4) the Euler--Lagrange equation gives
\[
\nabla_\mu({\cal L}_{\cal F}F^{\mu\nu})=0,~~~~~~~~~~~~~~~~~~~~~~~~~~~~~~~~~~~~~~~~~~~~~~~~~~~(A1)
\label{A1}
\]
where
\[
{\cal L}_{\cal F}=\frac{\partial {\cal L}}{\partial {\cal F}}=\frac{2\beta(\sigma-1){\cal F}-1}{4\pi(1+2\beta{\cal F})^{\sigma+1}}.~~~~~~~~~~~~~~~~~~~~~~~~~~~~~~~~~(A2)
\label{A2}
\]
The equation for the electric field, with spherical symmetry and Eq. (A1), becomes (${\cal F}=-E^2(r)/2$))
\[
\frac{1}{r}\frac{d(r^2E(r){\cal L}_{\cal F})}{dr}=0.~~~~~~~~~~~~~~~~~~~~~~~~~~~~~~~~~~~~~~~~~~~~~~(A3)
\label{A3}
\]
By virtue of Eq. (A2) and integrating Eq. (A3), we obtain
\[
\frac{E(r)(1+\beta(\sigma-1) E(r)^2)}{(1-\beta E(r)^2)^{\sigma+1}}=\frac{Q}{r^2},~~~~~~~~~~~~~~~~~~~~~~~~~~~~~(A4)
\label{A4}
\]
where $Q$ is the electric charge (the integration constant). At $\beta=0$, Eq. (A4) gives the Coulomb’s electric field $E_C(r)=Q/r^2$. It is convenient to define unitless variables
\[
x=\frac{r}{\beta^{1/4}\sqrt{Q}}, ~~~~~y=\sqrt{\beta}E.~~~~~~~~~~~~~~~~~~~~~~~~~~~~~~~~(A5)
\label{A5}
\]
Then Eq. (A4) becomes
\[
\frac{y(1+(\sigma-1)y^2)}{(1-y^2)^{\sigma+1}}=\frac{1}{x^2}.~~~~~~~~~~~~~~~~~~~~~~~~~~~~~~~~~~~~~~(A6)
\label{A6}
\]
From Eq. (A6), we obtain for small $x$ (and small $r$)
\[
y=1+{\cal O}(x).~~~~~~~~~~~~~~~~~~~~~~~~~~~~~~~~~~~~~~~~~~~~~~~~~~~~(A7)
\label{A7}
\]
Making use of Eqs. (A5) and (A7) one finds as $r\rightarrow 0$
\[
E(r)=\frac{1}{\sqrt{\beta}}+{\cal O}(r).~~~~~~~~~~~~~~~~~~~~~~~~~~~~~~~~~~~~~~~~~~~(A8)
\label{A8}
\]
As a result, we have the finite value of the electric field at the origin $E(0)=1/\sqrt{\beta}$ that is the maximum of the electric field.
The plot of $y$ versus $x$ is depicted in Fig. 6 for $\sigma=0.75,~1.5,~2$.
\begin{figure}[h]
\includegraphics [height=3.0in,width=4.0in]{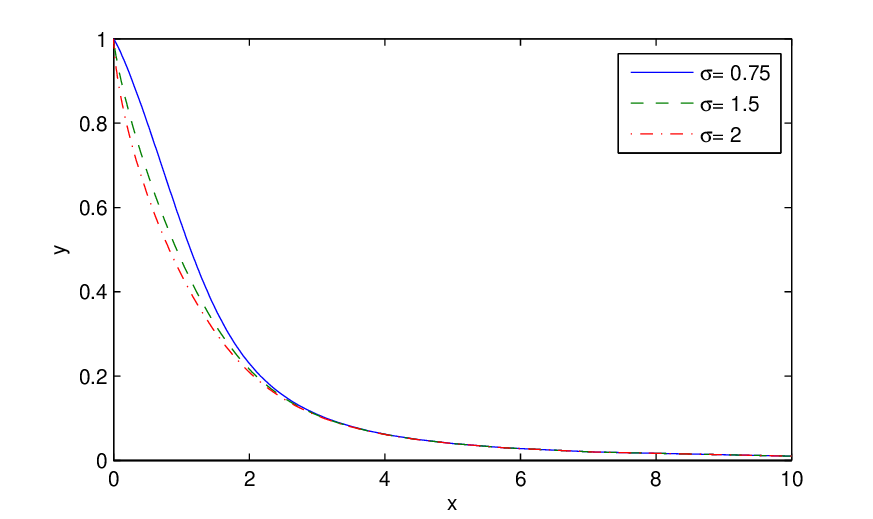}
\caption{\label{fig.6} The function $y$ vs. $x$ at $\sigma=0.75,~1.5,~2$.}
\end{figure}
We obtain from Eq. (A4) as $r\rightarrow \infty$
\[
E(r)=\frac{Q}{r^2}+{\cal O}(r^{-4}).~~~~~~~~~~~~~~~~~~~~~~~~~~~~~~~~~~~~~~~~(A9)
\label{A9}
\]
Equation (A9) shows that corrections to Coulomb's law are in the order of ${\cal O}(r^{-4})$.
According to Eq. (A7) and Fig. 6 the electric field is finite at the origin $r=0$ ($y=1$), and becomes zero as $r\rightarrow\infty$.
Because of nonlinearity of electric fields, an electric charge is not a real point-like object and does not possess a singularity at the center.\\

\textbf{Appendix B}\\

By virtue of Eq. (5) we obtain the electric energy density
\[
\rho=-E^2{\cal L}_{\cal F}-{\cal L}=\frac{E^2+\beta E^4(2\sigma-1)}{8\pi(1-\beta E^2)^{\sigma+1}}.~~~~~~~~~~~~~~(B1)
\label{B1}
\]
Making use of dimensionless variables (A5) one finds the electric energy density
\[
\rho=\frac{y^2+y^4(2\sigma-1)}{8\pi\beta(1-y^2)^{\sigma+1}}.~~~~~~~~~~~~~~~~~~~~~~~~~~~~~~~~~~~~~~~~~~(B2)
\label{B2}
\]
The total electric energy becomes
\[
{\cal E}=4\pi\int_0^\infty \rho(r)r^2dr=\frac{Q^{3/2}}{\beta^{1/4}}
\]
\[
\times\int_0^1\frac{(1-y^2)^{(\sigma-1)/2}[1+y^2(2\sigma-1)][(2\sigma^2-3\sigma+1)y^4+(5\sigma-2)y^2+1]dy}
{[1+(\sigma-1)y^2]^2\sqrt{y[y^2(\sigma-1)+1]}},~(B3)
\label{B3}
\]
where we  have used Eq. (A6). By numerical calculations of integral (B3) we obtain dimensionless variables $\bar{{\cal E}}\equiv {\cal E}\beta^{1/4}/Q^{3/2}$ which are presented in Table 1.
\begin{table}[ht]
\caption{Approximate values of $\bar{{\cal E}}\equiv {\cal E}\beta^{1/4}/Q^{3/2}$}
\centering
\begin{tabular}{c c c c c c c c c c c}\\[1ex]
\hline
$\sigma$ & 0.1 & 0.2 & 0.3  & 0.4 & 0.5 & 0.6 & 0.7 & 0.8 & 0.9 & 1\\[0.5ex]
\hline
$\bar{{\cal E}}$ & 1.272 &1.233 & 1.202 & 1.176 & 1.153 & 1.132 & 1.108 & 1.097 & 1.081 & 1.067\\[0.5ex]
\hline
\end{tabular}
\end{table}
As a result, in our NED model the electrostatic energy of charged objects is finite. According to the Abraham and Lorentz idea, the electron mass may be identified with the electromagnetic energy \cite{Born,Rohrlich,Spohn}. Then one can obtain the parameters $\beta^{1/4}$ and $\sigma$ to have the electron mass $m_e={\cal E}\approx 0.51$ MeV. Dirac also considered that the electron can be the classical charged object \cite{Dirac}.\\

\textbf{Appendix C}\\

We will obtain the solution with magnetic black holes.
The $tt$ component of Einstein's equation (3) with spherical symmetry (6) is given by
\[
f'(r)+\frac{f(r)}{r}=\frac{1}{r}+8\pi Gr\left(\rho_M(r)+\frac{\Lambda}{8\pi G}\right),~~~~~~~~~~~~~~~~~~~~~~(C1)
\label{C1}
\]
where
\[
\rho_M(r)=-T_t^{~t}=-{\cal L}(r).~~~~~~~~~~~~~~~~~~~~~~~~~~~~~~~~~~~~~~~~~~~~~~~~~(C2)
\label{C2}
\]
Equation (C1) is a linear first-order equation which belongs to the class of the general equation \cite{Boas}
\[
y'(x)+P(x)y(x)=Q(x),~~~~~~~~~~~~~~~~~~~~~~~~~~~~~~~~~~~~~~~~~~~~~~~(C3)
\label{C3}
\]
with the solution
\[
y(x)=\exp(-I(x))\int Q(x)\exp(I(x))dx+C\exp(-I(x)),~~~~~~~(C4)
\label{C4}
\]
where $C$ is the integration constant and
\[
I(x)=\int P(x)dx.~~~~~~~~~~~~~~~~~~~~~~~~~~~~~~~~~~~~~~~~~~~~~~~~~~~~~~~(C5)
\label{C5}
\]
Comparing Eqs. (C1) and (C3) one finds
\[
P(r)=\frac{1}{r},~~~Q(r)=\frac{1}{r}+8\pi Gr\left(\rho_M(r)+\frac{\Lambda}{8\pi G}\right),~~~~~~~~~~~~~~~(C6)
\label{C6}
\]
and $y(x)\rightarrow f(r)$, $x\rightarrow r$. Then from Eqs. (C4), (C5) and (C6) we obtain the solution to the $tt$ component of Einstein's Eq. (C1)
\[
f(r)=1+\frac{2Gm(r)}{r},~~~~~~~~~~~~~~~~~~~~~~~~~~~~~~~~~~~~~~~~~~~~~~~~~~~(C7)
 \label{C7}
\]
where we use the notations
\[
m(r)=m_0+4\pi\int r^2\rho_M(r)dr-\frac{r^3}{2Gl^2},~~~~C=2Gm_0,~~~~~~~~~(C8)
\label{C8}
\]
with $\Lambda=-3/l^2$.


\end{document}